\documentclass[journal]{IEEEtran}
\usepackage[lined,boxed]{algorithm2e}
\usepackage{cite}
\usepackage{times}
\usepackage{amsmath}
\usepackage{amssymb}
\usepackage{mathtools}
\usepackage{graphicx}
\usepackage[belowskip=5pt,aboveskip=5pt]{caption}
\usepackage{subcaption}
\usepackage{bm}
\usepackage{color}

\newtheorem{definition}{Definition}
\newtheorem{lemma}{Lemma}
\newtheorem{theorem}{Theorem}

\begin{document}

\title{Efficient Enumeration of  Unidirectional Cuts for Technology Mapping of Boolean Networks}

\author{Niranjan~Kulkarni,  Sarma~Vrudhula\\

School of Computing, Informatics and Decisions Systems Engineering\\
Arizona State University\\
Email: vrudhula@asu.edu \thanks{This research is supported by NSF PFI award no. 1237856 and NSF FRP award no. 1230401.}}%

\maketitle

\begin{abstract}
In technology mapping, enumeration of subcircuits or cuts to be replaced by a standard cell is an important step that decides both the quality of the solution and execution speed. In this work, we view cuts as set of edges instead of as set of nodes and based on it, provide a classification of cuts. It is shown that if enumeration is restricted to a subclass of cuts called unidirectional cuts, the quality of solution does not degrade. We also show that such cuts are equivalent to a known class of cuts called strong line cuts first proposed in\cite{kagaris1999maximum}. We propose an efficient enumeration method based on a novel graph pruning algorithm that utilizes network flow to approximate minimum strong line cut. The runtimes for the proposed enumeration method are shown to be quite practical for enumeration of a large number of cuts.
\end{abstract}

\section{Introduction}
\label{sec:introduction}

\IEEEPARstart{T}{echnology} mapping (TM) is a process of transforming a \emph{generic Boolean network}, which is a network consisting of primitive gates (e.g. AND/OR), into an equivalent \emph{mapped network}, that consists of  a network of \textit{cells} from a given technology library. Depending on the target implementation, the library cells can correspond to either lookup tables (LUT) in the case of an FPGA, or to a pre-designed set of standard cells in the case of an ASIC. The measures of delay, power, area, or some combination of them serve as an objective function to optimize during the transformation process. 

\begin{figure}[h]
\centering
\includegraphics[width=\columnwidth]{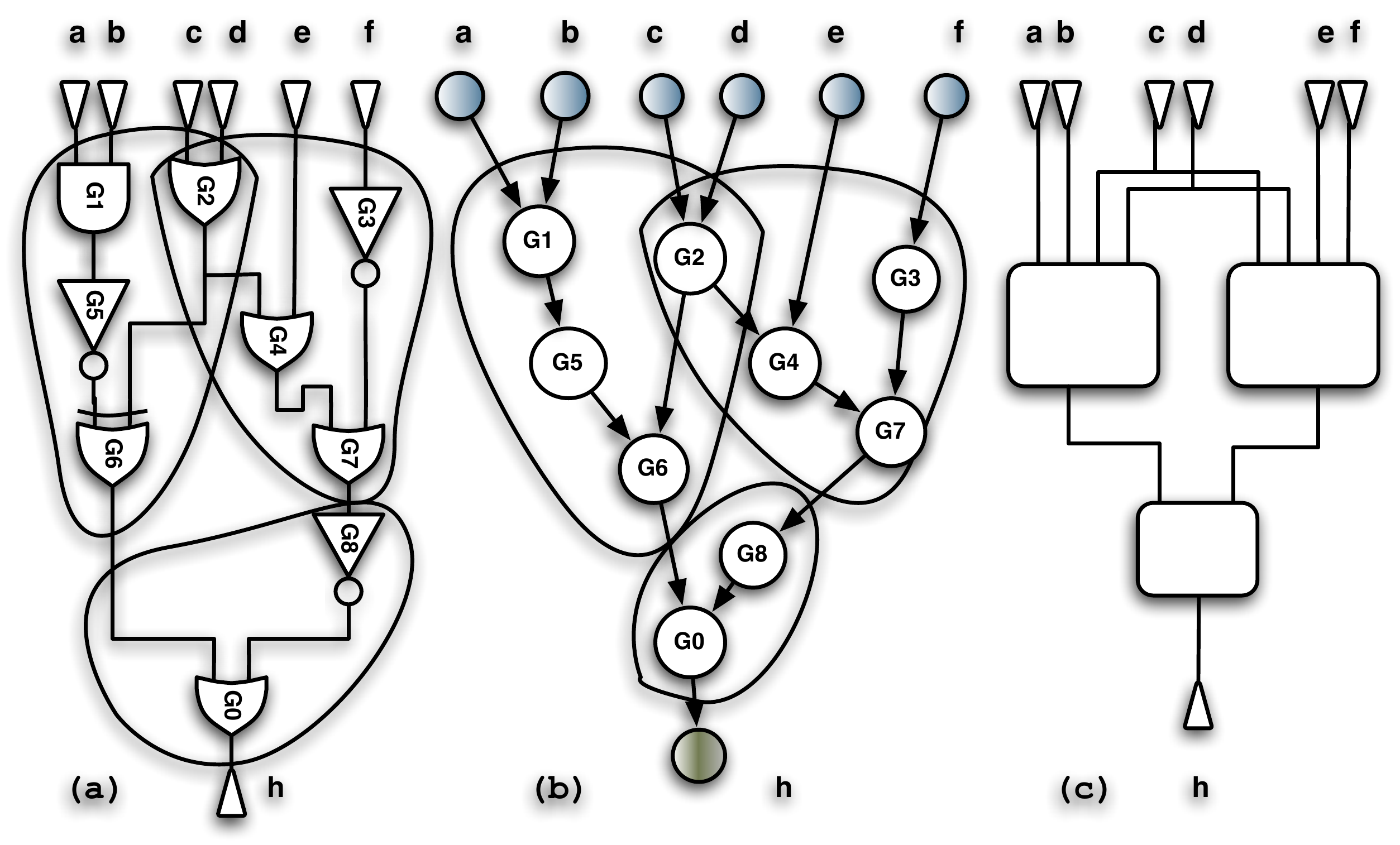}
\caption{a) A covered Boolean network. b) Its graph representation. c)   The network mapped on the library gates.}
\label{fig:covering}
\end{figure}

TM is formalized as a graph covering problem. A given Boolean network is represented as a \emph{directed acyclic graph} (DAG) $G = (V,E)$, which is referred to as the \emph{subject graph}.  The cells in the library, being single output Boolean functions, are also represented by DAGs, and each is referred to as a \textit{pattern graph}.  A feasible solution of TM is a complete covering of the subject graph with one or more of the pattern graphs (see Figure~\ref{fig:covering}). A core step in this process is to identify a subgraph in the subject graph to match with one or more pattern graphs. In the \emph{structural approach} to TM~\cite{chatterjee_thesis07}, the selection of subgraphs is achieved by computing a \emph{cut}.

The recent works on structural approaches to TM~\cite{Mishchenko2007,Chatterjee2006,Cong1999,Ling2007} are based on a particular type of a cut, called \emph{minimal node-cut}.  A \emph{node-cut} $C_v$ associated with a node $v \in V$ is a subset of nodes in the transitive fanin cone of $v$ such that every path from the primary inputs to $v$ includes a node in $C_v$. Note that in the existing literature, the definition of a node-cut restricts it to be \textit{minimal}, i.e., no proper superset of a node cut is a legal node cut. In this article we will explicitly refer to a minimal node cut, if that is the case. By definition, A \emph{$k$-feasible minimal node cut} of a node $v$ is a minimal node cut of cardinality $k$~\cite{Chatterjee2006}. 

A cut together with a node $v$ defines a single-sink subgraph that is either directly mappable onto a $k$-input LUT in the case of an FPGA,  or constitutes a match candidate for a NPN-equivalent standard cell, for an ASIC design, provided such an entity exists in the given library. In the literature $k$-feasible node cuts are usually simply referred as \emph{$k$-feasible cuts} as no other type of cut is usually considered for technology mapping \cite{Mishchenko2007,Chatterjee2006,Cong1999,Ling2007}.  

In Figure~\ref{fig:covering}, associated with node $G0$,  the sets $\{G6, G7\}$, and $\{G6, G8\}$ are 2-feasible minimal node cuts, whereas $\{G2, G5, G7\}$ is a 3-feasible minimal node cut. When the cut $\{G6, G7\}$ is selected, the actual subgraph replaced by a cell in TM consists of gates $\{G8,G0\}$.  Similarly, when the cut $\{G2, G5, G7\}$ is chosen the subcircuit consisting of gates $\{G8,G6,G0\}$ is replaced.

A node may have many associated $k$-feasible node cuts, which result in different coverings.  In TM, the \emph{quality} of a covering is usually  taken to be the  delay of the critical path, and/or the total area of the mapped circuit.  Hence, to evaluate the quality of the complete covering, cuts also need to be \emph{evaluated}. The quality of a cut is typically a combined measure of the subcircuit that will be replaced by a library cell and the subcircuit that feeds that cell. Since neither of these can be evaluated without knowing the cut, enumeration of cuts is a core task in TM.  Consequently, there has been a significant amount of effort devoted to the development of efficient algorithms for enumerating $k$-feasible node cuts~\cite{Chatterjee2006,Cong1994,Ling2007,Pan1998}.

While structural technology mapping turned out to be very successful, minimal node cuts bear some bias that eventually limits the number of possible matches. As demonstrated in \cite{Mishchenko_technologymapping}, this may result in excluding many feasible, high quality matches from the evaluation. The authors of \cite{Mishchenko_technologymapping} address this problem by constructing so-called \emph{supergates} i.e.\ single output networks of library gates that are added to the library in order to increase the number of matches. This demonstrates that enhancing a match space could yield significant benefits for structural technology mapping.

\begin{figure}[h]
\centering
\includegraphics[width=\columnwidth]{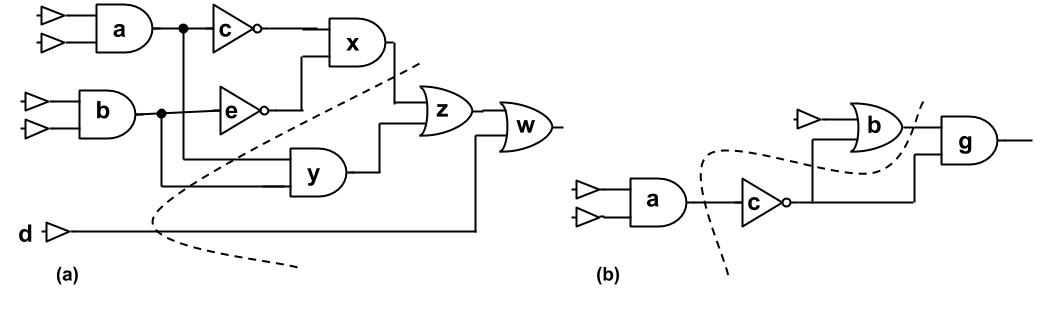}
\caption {(a) Unidirectional node cut denoted as $\{a,b,x,d\}$ (b) Bidirectional node cut denoted by $\{a,b\}$ }
\label{fig:bi-uni-example}
\end{figure}

Existing work on TM focuses on minimal node cuts. However, including non-minimal node cuts can increase substantially increase the possible matches.  Consider Figure~\ref{fig:bi-uni-example}.  The node cut $C = \{a,b,x,d\}$ is not minimal since $\{a,b,d\}$ is also a (minimal) node cut.  However $C$ corresponds to the function $ab + x + d$.  This happens to be a linearly separable (threshold) function, which would never be found by any cut enumerator that enumerates only minimal node cuts. Another representation of a cut, based on edges, is called a \textit{line cut}, which includes both minimal and non-minimal node cuts. 

\begin{definition}
\begin{enumerate}
\item[(a)] A line cut is a \textit{minimal} set of directed edges in a single sink DAG which when removed, eliminates all paths from any of source (primary input) to the sink i.e. produces an ``S-T bipartition''.   
\item[(b)] A line cut is \textit{k-feasible} if its cardinality (number of edges) is $k$ or smaller.
\item[(c)] A line cut is called a \textit{strong} line cut~\cite{kagaris1999maximum}  if no two edges in the line cut are on the same directed path. 
\end{enumerate}
\end{definition}

Note  that  line cuts and node cuts are simply two representations of same entity, and either form can be converted into other. They both partition the nodes of the DAG into two mutually exclusive sets $S$ and $T$. In such an S-T partition, the set $S$ of nodes must contain primary inputs and  $T$ must contain the sink node $v$. 

Figure~\ref{fig:bi-uni-example} a) shows a line cut consisting of edges $\{(x,z),(a,y),(b,y),(d,w)\}$. The node cut corresponding to this line cut is $C = \{a,b,x,d\}$, which is not minimal. Since minimal and non-minimal node cuts are useful, it appears that line cuts are the cuts that should be enumerated.  A cut can also be classified as being unidirectional or bidirectional. It is unidirectional if all the edges between $S$ and $T$ originate in $S$ and terminate in $T$. Otherwise it is bidirectional. 

In Figure~\ref{fig:bi-uni-example}(a) the line cut $\{(x,z),(a,y),(b,y),(d,w)\}$ (corresponding node cut being $C = \{a,b,x,d\}$) generates $S =\{a,b,c,d,e,x\}$, and $T = \{y,z,w\}$, and is unidirectional. In Figure~\ref{fig:bi-uni-example}(b), the line cut $\{(a,c),(b,g)\}$ (corresponding node cut is $\{a,b\}$) has $S = \{a,b\}$ and $T = \{c,g\}$, and is bidirectional since $(b,g)$ is a ``forward edge'' and $(c,b)$ is a ``backward edge''.  Note also that the minimal  node cut $\{a,b,d\}$ would identify $ab + a'b' + d$ as a function to be replaced by some cell from the library. However this is neither a member of any well defined class of functions, e.g. threshold functions, nor it is a function that would be in a typical cell library.  Thus, minimal cuts are not always the most useful or desirable. In addition, 
we show that bidirectional cuts are not necessary and discarding them does not degrade quality of TM with regard to critical path delay.  
Thus we need only enumerate  unidirectional node cuts (minimal or non-minimal). 

\begin{figure}[h]
\centering
\includegraphics[width=\columnwidth]{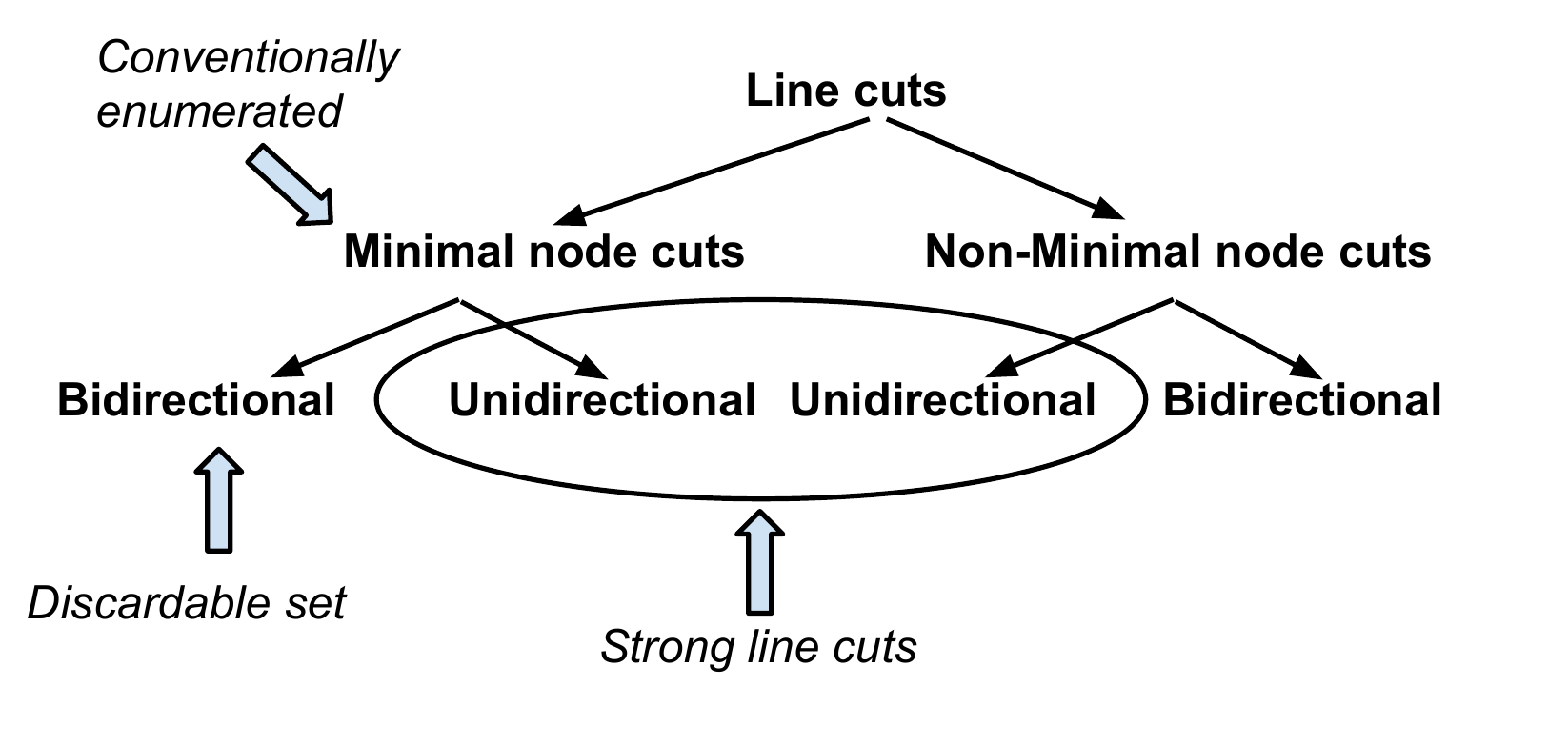}
\caption{Classification of cuts and their relationships}
\label{fig:cutrelations}
\end{figure}

We establish a one-to-one correspondence between unidirectional nod cuts and \textit{strong} line cuts~\cite{kagaris1999maximum}. This correspondence is important because there exists a well established relation between strong line cuts in a DAG and independent sets in its corresponding \emph{line dependency graph} (LDG)~\cite{kagaris1999maximum}. This allows for construction of efficient strong line cut enumeration techniques based on enumeration of independent sets. Since the latter is a problem well researched in computational graph theory, techniques exist that can be directly applied to enumerating line cuts by enumerating independent sets. Figure~\ref{fig:cutrelations} shows classification of cuts and their relationships.
The proposed technique for enumerating strong line cuts was used in \cite{Niranjan2010} to find threshold functions in a logic circuit.

The  main contributions of this article include:
\begin{itemize}
\item Introduction of unidirectional node cuts and a proof showing that restricting the cut space to only unidirectional node cuts does not degrade the quality of mapping for delay.
\item Establishing the equivalence unidirectional node cuts and strong line cuts.
\item An efficient $k$-feasible strong line cut enumeration algorithm based on the relationship between a DAG and its LDG. 
\item A general framework and the specific implementation of a pruning technique for $k$-feasible strong line cut enumeration.
\item An efficient implicit representation of bounded size MISs of a graph that allows for both unconstrained and constrained enumeration of such MISs.
\end{itemize}

\section{Related Work}
\label{sec:related-work}

The importance of cut computation in TM for FPGAs was first identified by the Cong \emph{et al.}~\cite{Cong1994}. They developed a novel and elegant network flow based algorithm that directly identified a single, depth-optimal, $k$-feasible node cut, without enumerating cuts.  Later, Pan \emph{et al.}~\cite{pan_lin,Pan1998} developed an efficient algorithm for enumerating cuts that avoided the large computational requirements of network flow.  More recently, Ling \emph{et. al}~\cite{Ling2007} developed a novel scheme for implicitly encoding cuts using Binary Decision Diagrams (BDD).  This representation allowed for extraction of cuts when the value of a cut could be computed recursively.  However, the authors admit that BDD approach is not very well suited for cut enumeration since non-cuts, which dominate cuts, are also implicitly included and need to be pruned during enumeration.

Finding a computationally efficient way of encoding and enumerating cuts is of fundamental importance to technology mapping. Recently Takata \emph{et al.\/}~\cite{takata2009efficient} proposed a top-down scheme for the procedure of~\cite{Pan1998} and demonstrated speed-ups of 3x-8x for larger $k=8,9$. Unfortunately, since the number of cuts of size at most $k$ is of $O(n^k)$, cut enumeration algorithms inherently suffer from poor scalability. To alleviate this problem, techniques for ranking and pruning of cuts were first proposed by Cong \emph{et al.}\ in \cite{Cong1999}. The basic observation of this work is that for certain optimization objectives it is possible to narrow the search down efficiently and extract depth-maximal or area-minimal cuts directly. Similar ideas, referred to as \emph{priority cuts}, were proposed by Mischenko \emph{et al.}\ in \cite{Mishchenko2007}, where appropriate seeding of the procedure from \cite{Pan1998} assured enumeration of only $O(n^2)$ priority cuts instead of $O(n^k)$ cuts.  These can be further  sorted by quality, and pruned. An alternative approach to pruning was proposed by Chatterjee \emph{et al.}\ in \cite{Chatterjee2006} where they introduced hierarchical partitioning of the cut space based on a novel concept that is similar to algebraic factorization. The authors showed that while complete factorization may still suffer from poor scalability, partial factorization of the cut space could yield good, practical solutions with very short runtimes.  Takata \cite{takata2009efficient} proposed a partial enumeration scheme that enumerates only a subset called \emph{label cuts}. The scheme improves scalability of cut enumeration and guarantees to maintain the circuit's depth, at the expense of small increase in the estimated network area. 

All the works identified above, and many others have demonstrated that \emph{structural technology mapping}, the core of which involves cut enumeration, leads to far superior solutions than the traditional graph/tree matching based algorithms. Cut enumeration has also found uses in related applications such as re-synthesis through rewriting \cite{Mishchenko2009}, application specific instruction set extension generation/optimization \cite{cong2006architecture}, hardware/software co-design \cite{peddersen2005rapid}, model checking in verification \cite{case2008cut}, and SAT problem preprocessing for simplification \cite{een2007applying}.

The use of a line dependency graph (LDG) derived from a DAG was proposed by Kagaris \emph{et al.}~\cite{kagaris1999maximum} to compute the maximum strong cut in a circuit for the purpose of delay testing. Based on the observation that an LDG is a transitively-oriented graph, hence a \emph{comparability graph}~\cite{golumbic2004algorithmic}, they provide an efficient and elegant algorithm that computes a maximum independent set of the LDG using network flow. This set represents a maximum strong cut in the corresponding DAG. While their approach generated interest in the area of delay-testing, we will demonstrate that there is still greater opportunity for further exploration and exploitation of the DAG/LDG duality for strong cut enumeration.

\section{Strong line cuts}

We now describe the relation between DAGs and their corresponding line dependency graphs (LDG).  An LDG is an undirected graph derived from a DAG that encodes the \emph{transitive dependencies} between DAG edges \cite{kagaris1999maximum}. Each edge $e$ of the DAG has a corresponding node $v$ in the LDG. Two nodes of an LDG are connected if and only if their corresponding lines in the DAG are \emph{transitively dependent}, i.e., there exists a path in the DAG from any source to any sink that contains both edges. Consequently, if an edge in DAG is transitively dependent on another edge, by definition the corresponding nodes in LDG will be neighbors. Since LDGs are by definition \emph{transitively oriented}, they are also \emph{comparability graphs}~\cite{golumbic2004algorithmic}.

An \emph{independent set} (IS) in a graph $G(V,E)$ is a set $S$ of vertices no two of which share an edge. A \emph{maximal independent
  set} (MIS) is an independent set which is not a proper subset of any other independent set in the graph. 

%
%
\begin{lemma}
\label{lem:mis}
(From \cite{kagaris1999maximum}) A strong line cut of a DAG forms a maximal independent set (MIS) in its
corresponding LDG.
\end{lemma}

\begin{figure}[h]
\centering
\includegraphics[width=0.8\columnwidth]{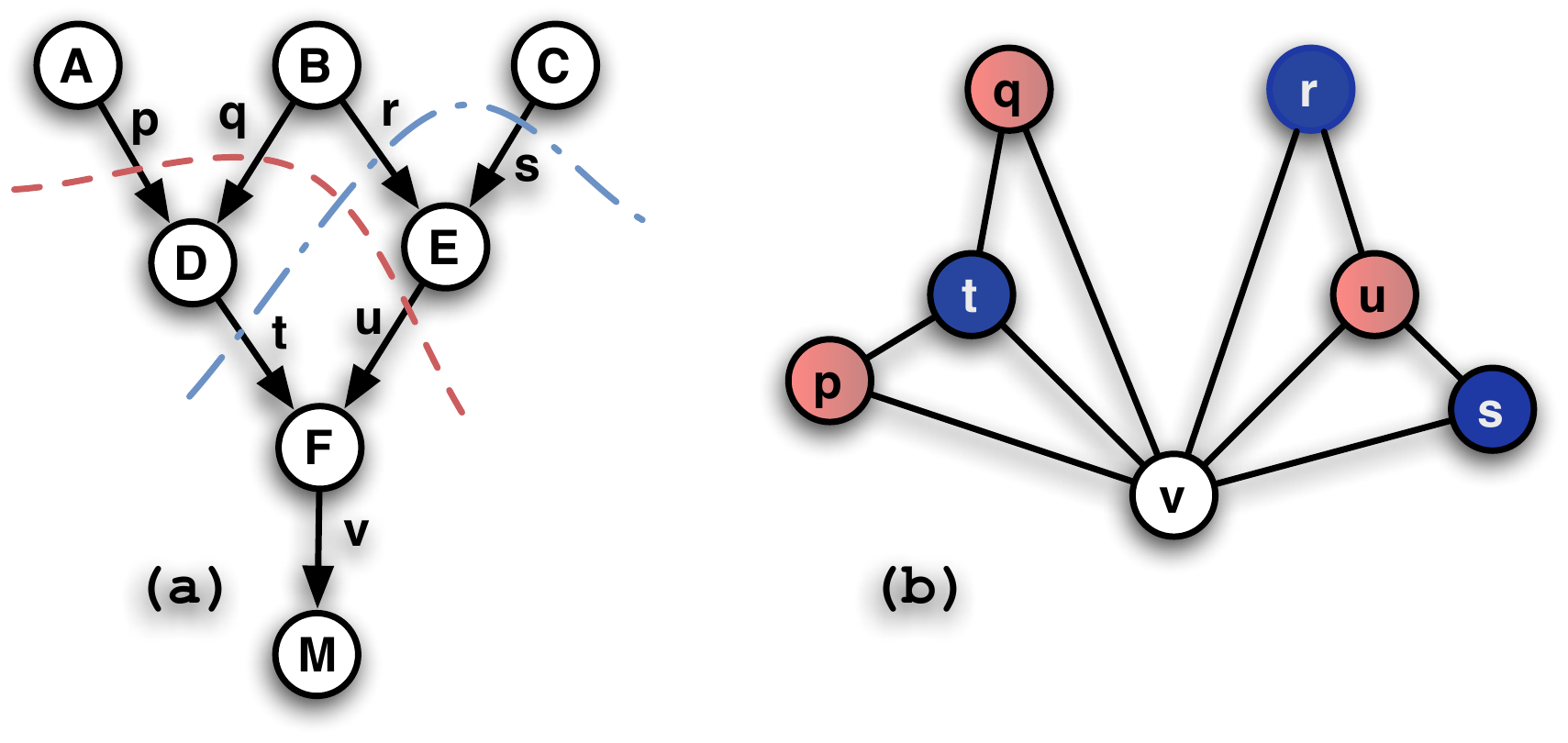}
\caption{\label{fig:cut} a) a DAG with strong cuts annotated. b) The corresponding  maximal independent sets in LDG.}
\end{figure}

Fig. \ref{fig:cut} illustrates the relation between DAGs and LDGs established by Lemma \ref{lem:mis}, on an example that we will use throughout this article  for illustration. The direct consequence of this lemma is that enumerating all $k$-feasible strong line cuts in a DAG is equivalent to enumerating all maximal independent sets of size $\leq k$ in the LDG.

\begin{figure*}[t]
\centering
\includegraphics[scale = 0.25]{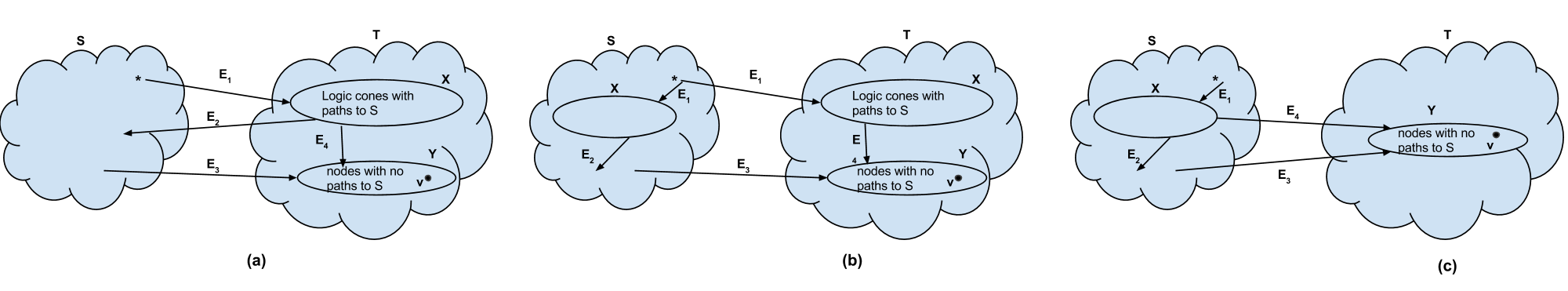}
\caption{\label{fig:bi-uni-cut-classification} a) Classification of edges in a bidirectional cut b) Replication after TM, (c) Classification of edges in corresponding unidirectional cut.}
\end{figure*}

\subsection{Relationship between unidirectional node cuts and strong line cuts}
In this section, we show the equivalence between unidirectional node cuts and strong line cuts. We also establish the fact that if the cut space is restricted to unidirectional node cuts then the quality of technology mapping for minimizing delay remains same.

In the following we restrict the DAG to be a transitive fan-in cone of some gate in a circuit, since in TM  only transitive fan-in cones of gates/nodes are considered for enumerating cuts. $v$ refers to the root node whose transitive fan-in cone is being considered. 

\begin{lemma}
\label{lem:unidirmis}
A strong line cut corresponds to a unidirectional set of edges crossing an $S-T$ partition.
\end{lemma}
\begin{IEEEproof}
For an arbitrary node $u \in S$, there exists a path $u \leadsto v$.  This is straightforward from the definition of DAG considered here which is the transitive fan-in cone of the node $v$. 
Assume that the $S-T$ partition corresponding to a strong line cut $C_v$ is bidirectional, i.e., there exists a directed edge $(p,q)$ such that $p \in T$ and $q \in S$. Then for some $x \in S$ and $y \in T$, $(x,y) \in C_v$,  there must exist a path $x \rightarrow y \leadsto p$. 
Since edge $(p,q)$ exists, there must exist a path $x \rightarrow y \leadsto p \rightarrow q$. Since $q \in S$, the root node $v$ must be reachable from $q$ through another edge in the cut $C_v$, say $(r,s)$. Therefore we have a complete path that looks like $x \rightarrow y \leadsto p \rightarrow q \leadsto r \rightarrow s \leadsto v$. Note that edges $(x,y)$ and $(r,s)$ both belong to the cut. This is clearly a contradiction, since no two lines in the cut $C_v$ should lie on same path. Also $(x,y) \neq (r,s)$ because that would lead to a directed cycle $x \rightarrow y \leadsto p \rightarrow q \leadsto x$ in the directed acyclic graph under consideration.

Conversely, assume a unidirectional node cut in which all the edges are from $S$ to $T$, and suppose the corresponding line cut is not a strong line cut.  Then there must exist at least edges $e_1 = (x,y)$ and $e_2 = (u,w)$ in the line cut that are on the same path to node $v$ (the output).  Assume $e_1$ precedes $e_2$ in the path.  By definition of a $S-T$ partition, $x \in S$, $y \in T$, $u \in S$ and $w \in T$.  However, since $y \leadsto u$, we have an edge starting from $T$ and ending in $S$, which contradicts the assumption that it is a unidirectional node cut. 
\end{IEEEproof}

Lemma~\ref{lem:unidirmis} confirms that a strong line cut must be unidirectional and a unidirectional cut must be a strong line cut. 
Note that the cardinality of a strong line cut and the unidirectional node cut can be different. The reason we convert back from a strong line cut to a node cut (which is unidirectional) is that eventually a node cut is what is mapped onto a cell. A node cut form of a line cut would always require smaller library cell whether mapping is done using standard Boolean functions or threshold functions.

Next we show that restricting node cuts to unidirectional node cuts will not increase the critical path delay when that is the objective function being minimized by TM.  Note that in TM, the delay of  path is the sum of the delays of gates in the path. We show that the set of paths to the output node $v$  in a bidirectional cut is the same as those in the corresponding unidirectional cut. 

Figure~\ref{fig:bi-uni-cut-classification}(a) shows a classification of the edges in a bidirectional cut.  $T = X \cup Y$, where $X$ is the set of logic cones (node and all nodes in its fanin cone) whose output has a directed edge to some node in $S$, and $Y$ is the set of nodes with no paths to $S$.  Note $v \in Y$.  TM would replicate $X$ in $S$ and then replace $T$ with some appropriate cell in the library.  This is depicted in Figure~\ref{fig:bi-uni-cut-classification}(b).  Four types of edges can be identified in the $S-T$ partition: (1)~$E_1$ are edges from $S$ to $X$, (2)~$E_2$ are edges from $X$ to $S$, (3)~$E_3$ are edges from $S$ to $Y$, and (4)~$E_4$ are edges from $X$ to $Y$. 

Now a path from input node in $S$ to the output node $v \in T$ can be one three types: 
\begin{enumerate}
\item $\textcircled{S} \stackrel {E_1}{\Longrightarrow} X \stackrel {E_4}{\Longrightarrow} Y \Longrightarrow v$.
\item $\textcircled{S} \stackrel {E_2}{\Longrightarrow} S \stackrel {E_3}{\Longrightarrow} Y \Longrightarrow v$. 
\item $\textcircled{S} \stackrel {E_3}{\Longrightarrow} Y \Longrightarrow v$. 
\end{enumerate}
Note that every one of the above paths (sequence of nodes) in the graph of Figure~\ref{fig:bi-uni-cut-classification}(a) also exists in the graph shown in Figure~\ref{fig:bi-uni-cut-classification}(b). Now consider the corresponding unidirectional cut shown in Figure~\ref{fig:bi-uni-cut-classification}(c). Every path that exists in Figure~\ref{fig:bi-uni-cut-classification}(a) also exists in Figure~\ref{fig:bi-uni-cut-classification}(c), and visa versa. This shows that there is no disadvantage to retaining only unidirectional cuts. 


\section{Cut enumeration}

Enumerating MISs is a key step in many computational graph theory problems~\cite{eppstein2001small,byskov2003algorithms,byskov2004enumerating}.  In TM, because there is a fixed library of functions, MIS enumeration needs to be restricted to sets of size $\leq k$.  Without any restrictions, the number of MISs in arbitrary graphs grows exponentially in the size of the graph~\cite{byskov2004enumerating}.  However, in TM, the size $k$ of the MIS is bounded above by some constant, and independent of $n$, which is the size of the graph. Therefore the number of MISs of size $\leq k$ is $\leq n^k$.  A brute force approach in which all subsets of size $\leq k$ are examined and those that are not an MIS are discarded, is not practical for even realistic values of $n$ and $k$. Existing algorithms exploit specific properties of small MISs to facilitate enumeration~\cite{eppstein2001small}. We now describe a method that can significantly speedup existing MIS enumeration algorithms by  pruning away many MISs that will not be part of the final solution. 

\subsection{MIS pruning}
The LDG of a DAG encodes MISs, many of which have sizes $> k$.  The basic idea in the pruning algorithm is to (efficiently) transform an LDG into a new, smaller, and denser graph $G'$ which contains all the MISs of size $\leq k$ of the original LDG, and as few other (parasitic) MISs as possible.  The objective is to construct a transformation which is computationally efficient and would significantly reduce the runtime of enumeration.

The graph $G'$ to be constructed must satisfy the  following conditions: every vertex $v$ of $G'$ as well as every disconnected pair of vertices in $G'$ must independently be a part of some MIS of size $\leq k$ of the original graph $G$. This condition  translates into two steps of the pruning algorithm. In the first step, for each vertex $v$ we attempt to determine the size of the smallest MIS to which $v$ belongs. If this MIS is of size $\leq k$ then $v$ is included in $G'$. The second step decides if any two disconnected vertices in $G$ may \textit{safely} share an edge in $G'$, implying that they will not be part of any MIS.  Again for each pair of disconnected vertices $(u, v)$ we attempt to determine the size of the smallest MIS containing both of the vertices. If such MIS is of size $> k$ then an edge $(u, v)$ is added to $G'$. This is the key step in the  of following pruning algorithm.

\begin{algorithm} 
  \KwIn{A DAG $D(V_N,E_N)$, An LDG $G = (V,E)$ and an integer $k$} \KwOut{ Graph $G' = (V',E')$ characterized above} \BlankLine $dl = \Phi$, $el = \Phi$\; \For{vertex $v$ in $G$} { $\lambda = $ Min-MIS($D, G, v$)\; \If{$\lambda > k$} { $dl = dl~\bigcup~v$\; } }

\For{disconnected pair $(u,v)$ in $G$ such that $u \notin dl$ and $v \notin dl$}
{
$\lambda = $ Min-MIS($D, G, u, v$)\;
\If{$\lambda > k$}
{
$el = el ~\bigcup~(u,v)$\;
}
}

$E' = E ~\bigcup~ el$\;
$V' = V~-~dl$\;
\Return $G'(V',E')$\;
\caption{Algorithm to prune MISs of an LDG}
\label{algo:prune} 
\end{algorithm} 

There is no known polynomial procedure to compute the size of the smallest MIS that contains a given vertex or a pair of vertices for comparability graphs \cite{corneil1984clustering}. Hence we \emph{approximate} the size of the minimum MIS containing a vertex $v$ or a pair of vertices ($u,v$) of a given LDG by exploiting the duality between MISs in LDGs and strong line cuts in DAGs. The minimum MIS in an LDG is the minimum strong cut in the DAG. We use this fact to approximate the minimum MIS size in an LDG by means of a flow computation in its corresponding DAG.

It is well known that a minimum $s$-$t$ cut (min-cut) is equivalent to the maximum flow the sink $t$ receives from the source $s$ \cite{cormen2001introduction,ford1962flow}. The size of a cut is a sum of capacities of edges involved in it. If unit edge capacities are assigned, then the size of a cut is equivalent to number of edges.  We note that an edge with a capacity of $\infty$ can never be part of a finite size cut.  The size of the minimum MIS in an LDG containing a vertex $v$ or a pair of vertices $(u,v)$ is approximated by computing the min-cut in the DAG with unit edge capacities. The procedure (Alg.\ \ref{algo:minmis}) assigns a capacity of $\infty$ to the dependent lines of the given line (e.g.\ corresponding to node $v$) and a capacity of 1 to all other lines. The capacities of edges attached to $s$ and $t$ are always $\infty$. This is because a min-cut must consist of circuit lines only.  Finally it returns the size of the minimum $s$-$t$ cut of the network ($\lambda$).

\begin{algorithm} 
\KwIn{A DAG $D(V_N,E_N)$, LDG $G$ and  a line $v$} 
\KwOut{Approximate size of minimum MIS (strong cut) containing $v$}
\BlankLine 
\For{$u$ in $G$}
{
$capacity[u] = 1$\;
}
\For{neighbor $w$ of $v$ in $G$}
{
$capacity[w] = \infty$\;
}
\Return Min-cut($D$,$capacity$)\;
\caption{Min-MIS procedure for single edge in DAG}
\label{algo:minmis} 
\end{algorithm} 

\begin{lemma} 
\label{lem:mincondition} 
Let $S_{min}$ be the minimum strong cut containing line $v$. Let $\lambda$ be the size of a min-cut of the network with capacities modified based on line $v$. Then $\lambda \leq |S_{min}|$.  
\end{lemma}
\begin{IEEEproof}
The Min-MIS procedure never assigns a capacity of $\infty$ to any line $l \in S_{min}$. Thus $|S_{min}|$ is finite. It immediately follows that $S_{min}$ is an arbitrary $s$-$t$ cut, and cannot be smaller than the min-cut of the network. 
\end{IEEEproof}

Lemma~\ref{lem:mincondition} states that the size of the minimum strong cut is guaranteed to be greater than or equal to the size of a min-cut. Hence if the result of Min-MIS is $>k$ then the size of a minimum strong cut is also $>k$. Consequently, the vertex in $G$ for which Min-MIS was computed to be $> k$ can be safely discarded from $G'$, or an unconnected pair of vertices for which Min-MIS was computed can be connected in $G'$. 

\begin{figure}[h]
\centering
\includegraphics[width=0.75\columnwidth]{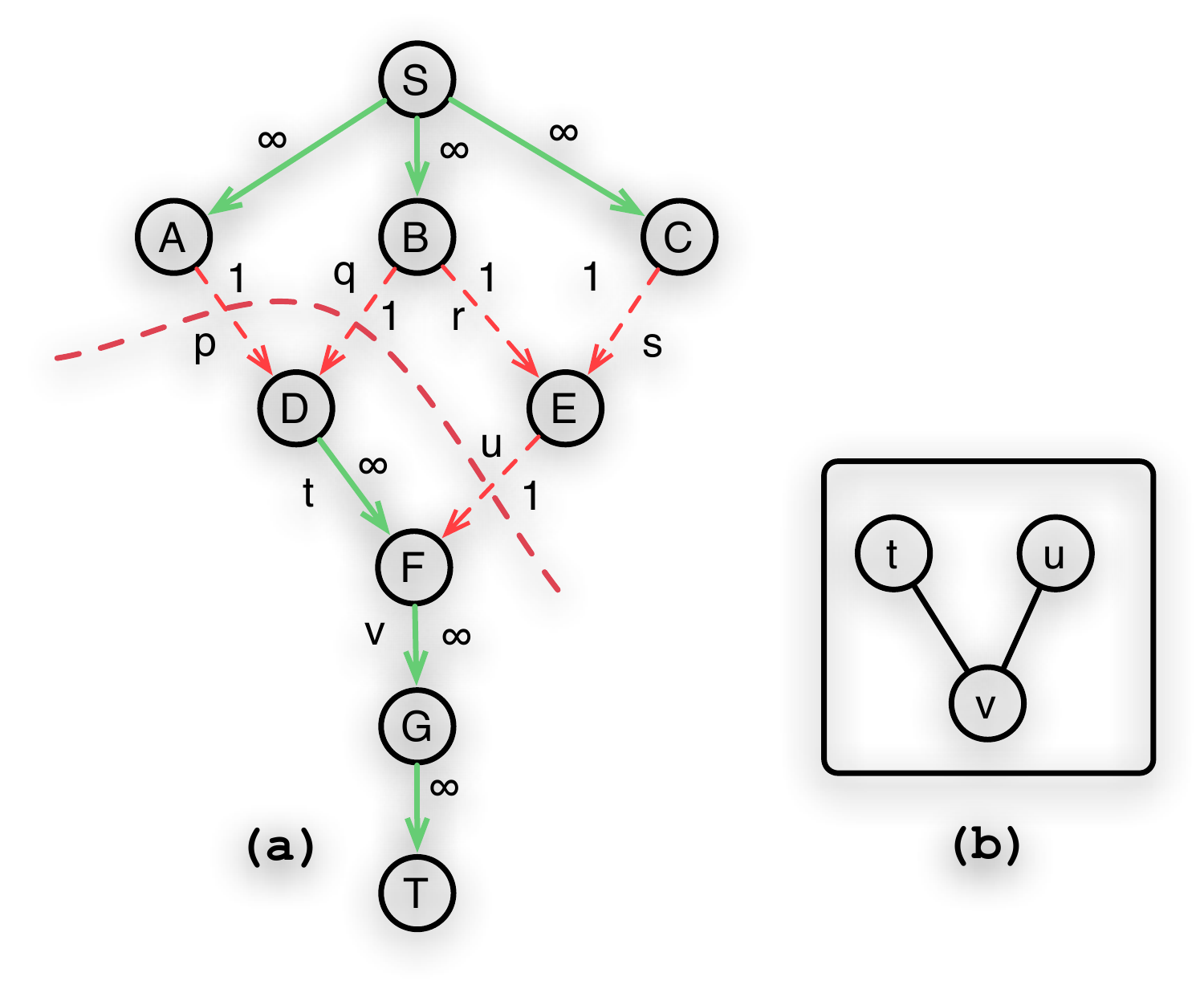}
\caption{a) Formation of $s-t$ Boolean network for determination of a  cut containing line $p$. b) LDG from Fig. \ref{fig:cut} pruned for
  $k \leq 2$.}
\label{fig:mincut}
\end{figure}

As an example, consider the  LDG in Fig.\ \ref{fig:cut}(b), and suppose we wish to check whether vertex $p$ belongs to an MIS of size $\leq 2$. This is equivalent to determining if line $p$ belongs to minimum strong cut of size at most 2 in the DAG.  We assign capacities to the edges, as shown in Fig.\ \ref{fig:mincut}(a).  Line $p$ is assigned a capacity of 1 and its dependent lines, $t$ and $v$, are assigned a capacity of $\infty$.  After this, the $s$-$t$ minimum cut size ($\lambda$) is determined. In this example, it is lines $p$, $q$ and $u$, and its size is 3 --- i.e. $\lambda = 3$.

\begin{theorem}
\label{thm:main}
Let $S$ be an MIS in $G$ such that $|S| \leq k$. Then $S$ is also an MIS in $G'$.
\end{theorem}
\begin{IEEEproof}
From Lemma \ref{lem:mincondition} every vertex of $S$ must be present in $G'$ and that no edge was added between any two of its vertices. Thus $S$ is an independent set in $G'$. Assuming $S$ is not a MIS in $G'$, there exists an independent set $S'$ in $G'$ such that $S \subset S'$. It follows that $S'$ is also an independent set in $G$ contradicting the fact that $S$ is an MIS in $G$.
\end{IEEEproof}

\begin{lemma}
Pruning algorithm runs in $O(n^3)$.
\end{lemma}
\begin{IEEEproof}
Let $n$ be the number of nodes, $m$ be the number of edges in DAG. The second \emph{for} loop in pruning algorithm runs in $O(m^2)$ and dominates the overall complexity. Determining the min-cut takes $O(km)$ time \cite{cormen2001introduction}. Since $k \leq n$ and is independent of $n$, the pruning has fixed complexity of $O(m^3)$. We know that $m \leq \Delta n$ where $\Delta$ is maximum degree found in the  DAG. For most of the circuits with limited fanin (and fanout) capacities $\Delta$ can be regarded as a small constant independent of $n$. Hence the time complexity of the pruning procedure is $O(n^3)$.
\end{IEEEproof}

As a result of the pruning transformation we perform MIS enumeration on $G'$ instead of $G$. Note that not all MISs in $G'$ are of size $k$ in $G$. However, our experiments demonstrate that using $G'$ instead of $G$ to enumerate MISs significantly reduces the enumeration time. In fact, the enumeration runtimes we observe in our experiments are practical for all of the evaluated benchmark circuits, suggesting that the approximation of a \emph{Min-MIS} used in pruning must be quite accurate.

\subsection{Enumerating MISs}

As stated earlier,  there are many known MIS enumeration techniques in computational graph theory.  In fact, the choice of enumeration algorithm is independent of  MISs pruning. In our experiments, we used a recursive procedure that is basically a simplified form of the algorithm presented in~\cite{byskov2003algorithms}. The idea is to recursively enumerate MISs that contain a specific node and then those that do not contain that same node. Once we designate a node to be part of the MIS, none of its neighbors can be part of that MIS. Similarly, if a node is not a part of the MIS, any of its neighbors can be part of the MIS. 


Note, that due to the pruning transformation being approximate, $G'$ may still contain some MISs of size $>k$. Since we are interested only in maximal independent sets of size $\leq k$, we simply discard, on the fly, the MISs whose size is greater than $k$.

Unfortunately, the number of MISs of a graph increases exponentially as a function of its size~\cite{moon1965cliques}. However in practice  we found that while running it on a pruned graph $G'$, enumeration time is dominated by pruning time for sufficiently small $k$.  Hence, even the simple recursive algorithm that we used is still efficient. More sophisticated approaches to enumerate MISs of size $\leq k$ \cite{byskov2003algorithms,eppstein2001small} could be used to improve runtime for with values of $k$. Their application on the transformed graph may further improve total runtimes as well as scalability of the solution.

\subsection{Results}

The procedures described in this article were implemented in C++ and run on a $2$GHz PC with 2 GB of RAM.  The results of these runs are summarized in Table \ref{tab:enumreduced}. We ran the simple cut enumeration algorithm after pruning the MISs, and enumerated \underline{all} cuts in the  ISCAS benchmark circuits. The starting point was an AND-INVERTER Graph (AIG) obtained from \cite{Mishchenko2007}. Note that MIS enumeration is not possible with any of the existing MIS enumeration schemes.  In fact, even a graph with as few as a hundred nodes would not be practical.  

The proposed pruning procedure makes it possible to exhaustively enumerate large numbers of strong line cuts in reasonable time. We also evaluated the combined effect of constraining the cone size and increasing the value of $k$.  The results demonstrate that for sufficiently small $k$, line cut enumeration is dominated by our pruning transformation and as such is practically polynomial. For larger values of $k$ however the procedure could however benefit from a more efficient MIS enumeration procedures. 

\begin{table*}[t]
\caption{Running times for enumeration}
\label{tab:enumreduced}
\def\thefootnote{a}\footnotesize
\centering
\begin{tabular}{|c|c|c|c|c|c|c|c|c|c|c|c|}
\cline{1-12}
& & & \multicolumn{3}{|c|}{$K = 6$, cone size = no limit}& \multicolumn{3}{|c|}{$K = 6$, cone size = 300}&\multicolumn{3}{|c|}{$K = 10$, cone size = 100}\\ \hline
{circuit} & {\# inputs} & {\# nodes} & {$\#$cuts} & pruning & enum.& {$\#$cuts} & {pruning} & enum. & {$\#$cuts} & {pruning} & enum.\\
             &                &              & after pruning& time (s) & time (s) & after pruning & time (s) & time (s) & after pruning & time (s) & time (s)\\ \hline
c432 & 36 & 356 & 4317 & 0.93 & 0.02 & 4253 & 0.98 & 0.02 & 382,697 & 2.22 & 1.60\\ \hline
c1355 & 41 & 1133 & 228,950 & 49.04 & 3.22  & 129,994 & 15 & 2.5 & 26,194,458 & 29.25 & 286.06\\ \hline
c1908 & 33 & 1793  & 9,519,182 & 48.12 & 9.43 & 2,099,375 & 18.49 & 5.02 & 1,311,442,759 & 36.46 & 6510.5 \\ \hline
c6288 & 32 & 4864  & 1,494,636 & 6085.06 & 190.86 & 737,587 & 480.83 & 63.96 & 150,921,850 & 379.75 & 4570.26 \\ \hline
c7552 & 207 & 7233  & 5,949,016 & 182.96 & 17.43 & 2,521,039 & 55.48 & 10.75 & 3,113,268,991 & 193.59 & 7046.68 \\ \hline
\end{tabular}
\end{table*}


\section{Conclusion}

In this work, we presented a novel cut enumeration framework that exploits duality between enumerable entities in DAGs and LDGs. Apart from resource efficient computational procedure, it also introduces into the area of technology mapping, the concept of $k$-feasible strong line cuts (or unidirectional cuts) that are distinct from conventional node cuts. The advantages are two-fold. On one hand they are enumerable with low per-unit computational effort. On the other hand they potentially open up a new space available to be explored by the technology mapper without degrading the quality of the mapping. Line cuts provide choices not available to node cut based technology mappers. More importantly line cuts inherently mitigate some of the structural bias of node cuts and unlike node cuts they guarantee a mappable cut of a circuit.

\bibliographystyle{abbrv}

\begin{thebibliography}{10}

\bibitem{byskov2003algorithms}
J.~Byskov.
\newblock {Algorithms for k-colouring and finding maximal independent sets}.
\newblock In {\em Proc. Symp. on Discrete Algorithms}, pages 456--457. PA, USA,
  2003.

\bibitem{byskov2004enumerating}
J.~M. Byskov.
\newblock Enumerating maximal independent sets with applications to graph
  colouring.
\newblock {\em Oper. Res. Lett.}, 32(6):547--556, Nov. 2004.

\bibitem{case2008cut}
M.~L. Case, A.~Mishchenko, and R.~K. Brayton.
\newblock {Cut-based inductive invariant computation}.
\newblock In {\em Proc. IWLS'08}, pages 172--179, 4--6 Jun. 2008.

\bibitem{Chatterjee2006}
S.~Chatterjee, A.~Mishchenko, and R.~K. Brayton.
\newblock Factor cuts.
\newblock In {\em Proc. ICCAD '06}, pages 143--150, 5--9 Nov. 2006.

\bibitem{Cong1994}
J.~Cong and Y.~Ding.
\newblock {FlowMap}: an optimal technology mapping algorithm for delay
  optimization in lookup-table based {FPGA} designs.
\newblock {\em IEEE Trans. CAD}, 13(1):1--12, Jan. 1994.

\bibitem{cong2006architecture}
J.~Cong, G.~Han, and Z.~Zhang.
\newblock {Architecture and compiler optimizations for data bandwidth
  improvement in configurable processors}.
\newblock {\em IEEE Trans. on VLSI Syst.}, 14(9):986--997, 2006.

\bibitem{Cong1999}
J.~Cong, C.~Wu, and Y.~Ding.
\newblock Cut ranking and pruning: enabling a general and efficient {FPGA}
  mapping solution.
\newblock In {\em Proc. FPGA'99}, pages 29--35, New York, 21--23, Feb. 1999.
  ACM.

\bibitem{cormen2001introduction}
T.~Cormen, C.~Leiserson, R.~Rivest, and C.~Stein.
\newblock {\em {Introduction to algorithms}}.
\newblock MIT Press, Cambridge, MA, 2001.

\bibitem{corneil1984clustering}
D.~Corneil and Y.~Perl.
\newblock {Clustering and domination in perfect graphs}.
\newblock {\em Discrete Applied Mathematics}, 9(1):27--39, 1984.

\bibitem{een2007applying}
N.~Een, A.~Mishchenko, and N.~Sorensson.
\newblock Applying logic synthesis for speeding up {SAT}.
\newblock {\em Lect. N. Comp. Sci.}, 4501:272, 2007.

\bibitem{eppstein2001small}
D.~Eppstein.
\newblock {Small maximal independent sets and faster exact graph coloring}.
\newblock {\em Lect. N. Comp. Sci.}, pages 462--470, 2001.

\bibitem{ford1962flow}
L.~Ford and D.~Fulkerson.
\newblock {Flow in networks}.
\newblock {\em Princeton University Press, Princeton, NJ}, 1962.

\bibitem{golumbic2004algorithmic}
M.~Golumbic.
\newblock {\em {Algorithmic graph theory and perfect graphs}}.
\newblock North-Holland, 2004.

\bibitem{kagaris1999maximum}
D.~Kagaris and S.~Tragoudas.
\newblock {Maximum weighted independent sets on transitive graphs and
  applications}.
\newblock {\em Integration, the VLSI Journal}, 27(1):77--86, 1999.

\bibitem{Niranjan2010}
N.~Kulkarni, N.~Nukala, and S.~Vrudhula.
\newblock Minimizing area and power of sequential cmos circuits using threshold
  decomposition.
\newblock In {\em Proceedings of the International Conference on Computer-Aided
  Design}, ICCAD '12, pages 605--612, New York, NY, USA, 2012. ACM.

\bibitem{Ling2007}
A.~C. Ling, J.~Zhu, and S.~D. Brown.
\newblock {BddCut}: Towards scalable symbolic cut enumeration.
\newblock In {\em Proc. ASP-DAC'07}, pages 408--413, 23--26 Jan. 2007.

\bibitem{Mishchenko2009}
A.~Mishchenko, R.~Brayton, J.-H.~R. Jiang, and S.~Jang.
\newblock Scalable don't-care-based logic optimization and resynthesis.
\newblock In {\em Proc. FPGA '09}, pages 151--160, NY, 21--23 Feb. 2009. ACM.

\bibitem{Mishchenko_technologymapping}
A.~Mishchenko, S.~Chatterjee, R.~Brayton, X.~Wang, and T.~Kam.
\newblock Technology mapping with boolean matching, supergates and choices.

\bibitem{Mishchenko2007}
A.~Mishchenko, S.~Cho, S.~Chatterjee, and R.~Brayton.
\newblock Combinational and sequential mapping with priority cuts.
\newblock In {\em Proc. ICCAD'07}, pages 354--361, 4--8 Nov. 2007.

\bibitem{moon1965cliques}
J.~Moon and L.~Moser.
\newblock {On cliques in graphs}.
\newblock {\em Israel Journal of Mathematics}, 3(1):23--28, 1965.

\bibitem{pan_lin}
P.~Pan and C.-C. Lin.
\newblock {A new retiming-based technology mapping algorithm for LUT-based
  FPGAs}.
\newblock In {\em Proc. FPGA '98}, pages 35--42, New York, 22--24 Feb. 1998.
  ACM.

\bibitem{Pan1998}
P.~Pan and C.~L. Liu.
\newblock Optimal clock period {FPGA} technology mapping for sequential
  circuits.
\newblock {\em ACM TODAES}, 3(3):437--462, 1998.

\bibitem{peddersen2005rapid}
J.~Peddersen, S.~Shee, A.~Janapsatya, and S.~Parameswaran.
\newblock {Rapid embedded hardware/software system generation}.
\newblock In {\em Int. Conf. on VLSI Design}, pages 111--116, 2005.

\bibitem{chatterjee_thesis07}
{S. Chatterjee}.
\newblock {On Algorithms for Technology Mapping, PhD Thesis}.
\newblock 2007.

\bibitem{takata2009efficient}
T.~Takata and Y.~Matsunaga.
\newblock An efficient cut enumeration for depth-optimum technology mapping for
  lut-based fpgas.
\newblock In {\em Proceedings of the 19th ACM Great Lakes Symposium on VLSI},
  GLSVLSI '09, pages 351--356, New York, NY, USA, 2009. ACM.

\end{thebibliography}

\end{document}